\pdfoutput=1
\documentclass[a4paper, conference]{IEEEtran}
\IEEEoverridecommandlockouts
\usepackage{xurl}
\usepackage{hyperref}
\usepackage{cite}
\usepackage{amsmath,amssymb,amsfonts}
\usepackage{graphicx}
\usepackage{textcomp}
\usepackage{xcolor}
\usepackage{graphicx}
\usepackage{caption}
\usepackage{subcaption}
\usepackage{float}
\usepackage{orcidlink}
\usepackage{stackengine}

\usepackage{fancyhdr}
\fancypagestyle{firstpage}
{
    \fancyhead[L]{\footnotesize \copyright~2026 IEEE. Personal use of this material is permitted. Permission from IEEE must be obtained for all other uses, in any current or future media, including reprinting/republishing this material for advertising or promotional purposes, creating new collective works, for resale or redistribution to servers or lists, or reuse of any copyrighted component of this work in other works. This paper is accepted at the 2026 IEEE International Symposium on Defect and Fault Tolerance in VLSI and Nanotechnology Systems (DFT).}
    \fancyhead[R]{}
}
\begin{document}

\title{JASPER: Special Session on \underline{J}oint Reliability \underline{A}nd \underline{S}ecurity Assessment of S\underline{P}lit Computing for \underline{E}dge \underline{R}obustness}

\author{
    \IEEEauthorblockN{
        E. Magliano\orcidlink{0009-0007-4089-1142}$^{1}$,
        G. Esposito\orcidlink{0009-0002-2850-0202}$^{1}$,
        A. Shahdadian\orcidlink{0009-0008-3761-4598}$^{2}$,
        R. Mounika Kodamanchili$^{3}$,
    }\\
    \IEEEauthorblockN{
        J.D. Guerrero Balaguera\orcidlink{0000-0001-6852-2372}$^{1}$,
        J.E. Rodriguez Condia\orcidlink{0000-0001-5957-5624}$^{1}$,
        A. Ruospo\orcidlink{0000-0003-2040-9762}$^{1}$,
        R. Siciliano\orcidlink{0000-0002-8012-0461}$^{2}$,
        S. Di Carlo\orcidlink{0000-0002-7512-5356}$^{1}$
    }\\
    \IEEEauthorblockN{
        M. Jenihhin\orcidlink{0000-0001-8165-9592}$^{3}$,
        M. Levorato\orcidlink{0000-0002-6920-4189}$^{4}$,
        A. Savino\orcidlink{0000-0003-0529-7950}$^{1}$,
        M. Sonza Reorda\orcidlink{0000-0003-2899-7669}$^{1}$,
        C. Herglotz $^{5}$,
        M. Hübner $^{6}$,
        M. Taheri\orcidlink{0000-0001-5405-992X} $^{6,3}$
    }\\
    \textsuperscript{1} \textit{Dept. of Control and Computer Engineering, Politecnico di Torino, Turin, Italy} \\
    \textsuperscript{2} \textit{Dept. of Electrical Engineering and Information Technology, University of Naples "Federico II", Naples, Italy} \\
    \textsuperscript{3} \textit{Dept. of Computer Systems, Tallinn University of Technology, Tallinn, EE} \\
    \textsuperscript{4} \textit{Computer Science Department, University of California, Irvine, US} \\
    \textsuperscript{5} \textit{Dept. of Computer Engineering, Brandenburg University of Technology, Cottbus, DE} \\
    \textsuperscript{6} \textit{Dept. of Computer Science, Chair of Computer Engineering, Humboldt University of Berlin, Berlin, Germany} \\

\bstctlcite{BSTcontrol}

\thanks{M. Taheri coordinated this Special Session. The main technical developments were carried out by E. Magliano, G. Esposito, and A. Shahdadian.}
}

\maketitle
\thispagestyle{firstpage}

\begin{abstract}

Split Computing (SC) enables efficient deployment of Deep Neural Networks (DNNs) by partitioning inference between edge devices and cloud servers. However, intermediate feature representations are simultaneously exposed to hardware faults and adversarial attacks, which are traditionally evaluated independently. This paper presents a unified framework for the joint assessment of reliability and security in Split Computing. First, reliability is characterized through neuron-level fault injection using the Mean Relative Accuracy Degradation (MRAD) while security through feature-map-aware adversarial attacks simulations using the Attack Success Rate (ASR). Based on these complementary analyses, the Joint Vulnerability Score (JVS) is introduced, along with a confidence-aware extension that jointly captures prediction errors and confidence degradation. The framework is evaluated on ten Split Computing configurations based on ResNet-50 trained on ILSVRC-2012. Experimental results show substantial differences across compression strategies, with MRAD ranging from 44.3\% to 61.2\% under fault injection, while adversarial attacks achieve up to 98.8\% ASR. Furthermore, the proposed joint metrics reveal vulnerability trends that remain hidden when reliability and security are analyzed independently, providing a more comprehensive methodology for designing dependable Split Computing systems.

\end{abstract}


\section{Introduction}
\label{sec:introduction}

Split Computing (SC) has emerged as a promising paradigm for executing Deep Neural Network (DNN) inference on resource-constrained edge devices by partitioning the DNN between an edge device (i.e., Head model) and a cloud server (i.e., Tail model). This approach combines the low-latency response and reduced energy consumption of edge computing with the computational capabilities of cloud infrastructures, while maintaining competitive inference accuracy \cite{matsubara2022split}. To further reduce communication overhead, intermediate feature maps exchanged between the edge and the server are commonly compressed using techniques such as channel reduction and quantization \cite{matsubara2023sc2}. These optimizations make SC an attractive solution for deploying large DNNs in safety-critical applications including autonomous driving, intelligent robotics, and industrial automation.

\begin{figure*}
    \centering
    \includegraphics[width=0.72\textwidth]{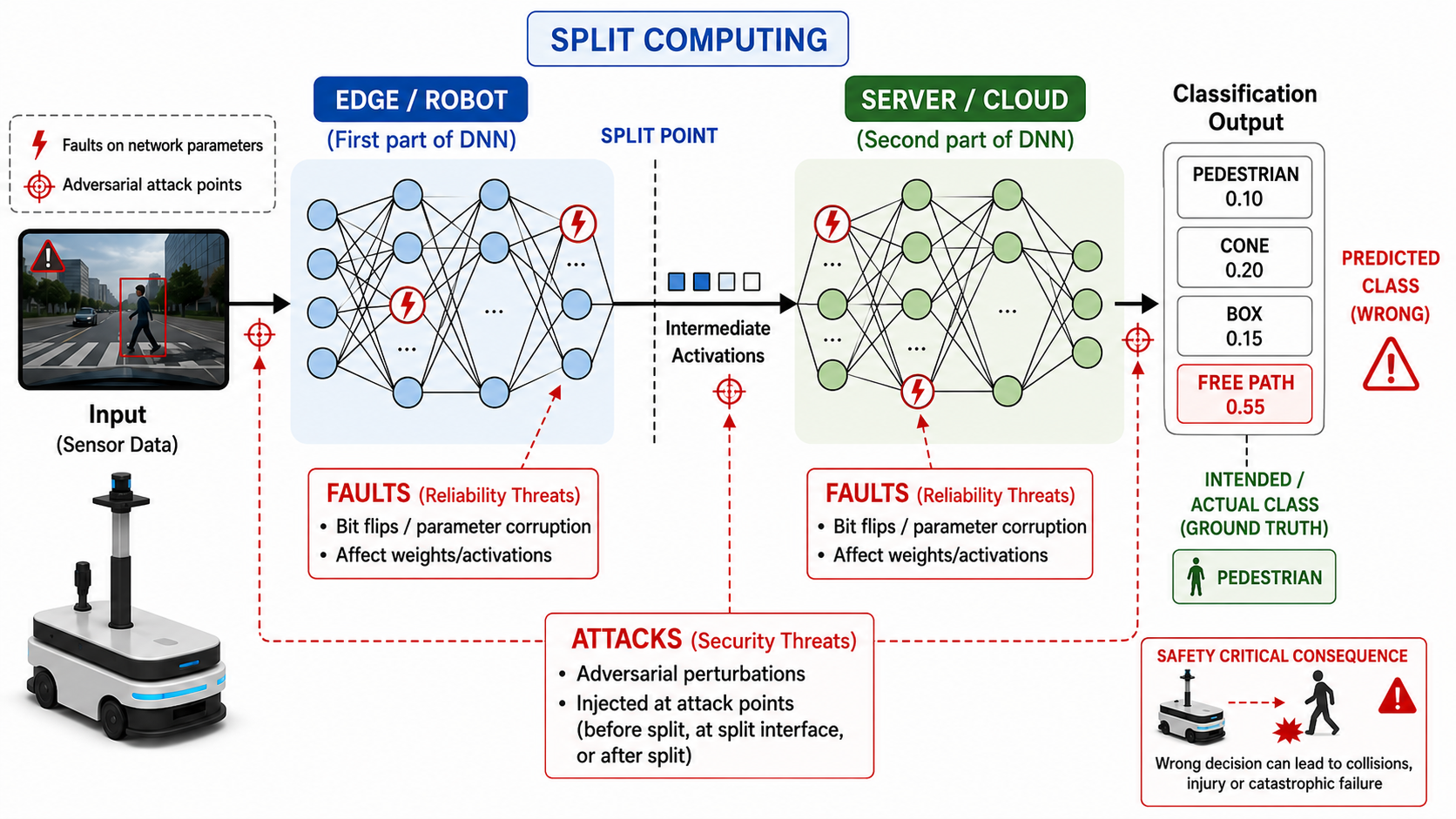}
    \caption{Overview of the reliability and security threats considered in SC and their impact on safety-critical inference.}
    \label{example}
\end{figure*}

Unlike conventional monolithic inference, however, SC introduces a distributed execution pipeline in which intermediate feature representations are transmitted through communication channels before the inference is completed. This additional interface significantly enlarges the attack surface of the system and exposes internal DNN representations to different classes of perturbations. On one hand, hardware faults originating from transient or permanent failures may corrupt DNN intermediate computations, leading to Silent Data Corruptions (SDCs) which, in turn, can lead to accuracy degradation and unreliable predictions \cite{hamdioui2013reliability, moore2020international, esposito2024assessing}. On the other hand, adversarial perturbations can deliberately manipulate intermediate feature maps to induce incorrect classifications \cite{goodfellow2015explaining,madry2017towards}. Although both phenomena ultimately manifest as incorrect decisions  (Fig. \ref{example}), they are traditionally investigated independently using different evaluation methodologies and robustness metrics.

This separation provides only a partial understanding of the overall vulnerability of SC systems. Optimizations designed to improve reliability may unintentionally affect adversarial robustness, while techniques introduced to enhance security may alter the propagation of hardware-induced errors \cite{taheri1, 7574542}. Consequently, evaluating these threats independently may lead to misleading conclusions regarding the overall dependability of the system, particularly in safety-critical scenarios where both accidental faults and malicious attacks may coexist \cite{esposito2026ai}.

To address this challenge, this work proposes a unified framework for the joint evaluation of reliability and security in SC. First, the reliability of different split configurations is analyzed by investigating the impact of neuron-level bit-flip faults under different feature-map compression strategies. A normalized reliability metric based on model robustness against fault injection is derived to characterize the susceptibility of each configuration. Second, the security perspective is investigated by evaluating the robustness of intermediate feature representations against adversarial perturbations, establishing a normalized security metric based on the attack success rate across different split configurations.

Building upon these complementary analyses, the paper introduces a Joint Vulnerability Score (JVS) that combines both reliability and security into a single quantitative metric, enabling a holistic assessment of SC robustness. Furthermore, the framework is extended by incorporating prediction confidence into the evaluation, allowing the proposed metric to capture not only misclassification events but also the degradation of the model's confidence under perturbations. 

All the experiments are performed using a ResNet-50 model trained on the ImageNet-2012 dataset under ten different SC configurations. Six configurations employ the Channel Reduction and Bit Quantization (CRBQ) compression strategy, namely CRBQ(1), CRBQ(2), CRBQ(3), CRBQ(6), CRBQ(9), and CRBQ(12). The remaining four configurations are based on the Entropic Student compression method with compression factors $\beta={0.64, 1.28, 2.56, 5.12}$. The evaluation dataset consists of 625 images selected from ImageNet-2012, representing the major super-classes within the original 1,000 ImageNet categories. This representative subset provides a diverse evaluation while maintaining a manageable experimental complexity for the extensive fault injection and adversarial attack campaigns performed in this work. \footnote{The code used to evaluate the SC models is available in this repository. \url{https://github.com/GiuseppeEsposito98/Anomaly-detection-and-Classification-in-SC2}}

The remainder of this paper is organized as follows. Section II presents the fault injection framework and reliability analysis of different SC configurations. Section III analyze the security of the considered configurations, while Section IV introduces the proposed Joint Vulnerability Score and its confidence-aware extension for the unified assessment of reliability and security. Finally, the work is concluded in Section V.

\section{Reliability}
\label{sec:reliability}

\subsection{Evaluation strategy}
\label{subsec:rel_eval_strat}

\begin{figure*}[!t]
    \centering
    \includegraphics[width=\textwidth]{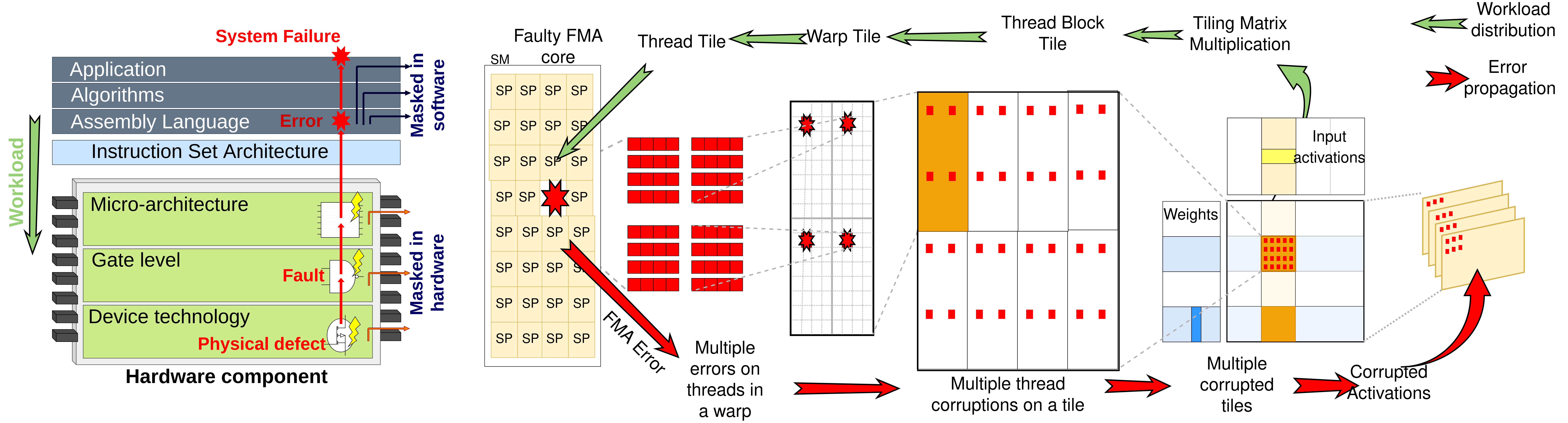}
    \caption{A general scheme from \protect\cite{esposito2025comparing} of physical defects propagation across system components (\textit{Left}), and an illustration of fault propagation effects on matrix multiplication operations through different levels of the tiling (\textit{Right}).}
    \label{fig:error_model}
\end{figure*}

    We focused our evaluations on hardware faults affecting mobile devices in SC architectures (e.g., drones or robots), since such devices use Commercial-Off-The-Shelf embedded Graphical Processing Units (GPUs) \cite{Levorato} that do not necessarily meet safety or high-reliability requirements \cite{amazonDrones}. 
    Thus, hardware faults in mobile devices may compromise the proper operation of an SC system. 
    It is worth noting that, for this work, our analysis focuses on computing errors on the mobile device while assuming ideal operation of the communication channels and of the cloud infrastructure executing the Tail model.

    The reliability evaluation was performed via fault injection (FI) campaigns that corrupt the \texttt{Head model} of the targeted SC DNN at the application level, propagating their effects through the SC DNN architecture during inference. 
    These FI campaigns introduce error perturbations that model fault effects from the underlying hardware accelerator in the mobile device. 

    We adopt a hardware-aware error model that mimics fault effects by corrupting the intermediate Feature Maps in the SC Head to model SDCs. 
    In particular, such SDCs represent effects from permanent faults or timing effects due to aging, or wear-out, inside defective Multiply-Accumulator (MAC) units of an embedded GPU device \cite{guerrero2024evaluating, esposito2025comparing}.

    For the evaluation, multi-bit flips are injected in the intermediate feature maps of the Head model, while following the tiling distribution and execution of matrix multiplications (MMs) as fundamental primitives for the deployed operations (i.e.,\,mostly\,Convolutions\,or\,Linear\,layers) on embedded GPUs.
    
    Figure \ref{fig:error_model} (\textit{right}) illustrates an example of the workload distribution as tiles on GPUs at different abstraction levels, as well as the impact of a fault during the execution of Tiling MM algorithms. 
    When one Streaming Multiprocessor (SM) contains a faulty MAC unit or Scalar Processor (SP), errors generated during MM tile execution can propagate to one or more threads, resulting in multiple data corruptions that affect the computed tile values.
    Indeed, similar tile corruption can occur when multiple MM tiles are processed by the same faulty SM.
    
    Considering the tile-based fault propagation, the tensor outputs from a targeted convolutional layer under evaluation are mapped to evaluate corruptions on tiling MM through three parameters: 
    \textit{i)} \textit{Tile Size} (TS) that corresponds to the tile size at device level, \textit{ii)} \textit{the Block Error Rate} (BER) that is the portion of tiles executed in an SM with a faulty MAC, and \textit{iii)} the \textit{Neuron Error Rate} (NER), which is the fraction of neurons within a tile corrupted due to a faulty MAC.
    Then, specific locations of the feature maps are corrupted by injecting random single-bit flips according to the coordinates defined by the selected tiles given TS, BER, and NER from the virtual tiling representation.

    The inference output obtained under faulty conditions is compared against the fault-free output to assess the impact of the injected errors from two complementary perspectives. First, we analyze the deviation in the SC DNN classification capabilities with respect to the golden execution. A change in the predicted class is generally regarded as one of the most critical consequences of a fault, since it directly affects the correctness of the system decision-making process \cite{rech2024artificial}. Therefore, we measured the \textit{Mean Relative Accuracy Degradation} (MRAD) \cite{hong2019terminal} which provides a direct measure of the capability of the SC DNN to preserve its intended functionality in the presence of hardware-induced perturbations.

    On the other hand, hardware faults may perturb the DNN logits without necessarily changing the predicted class, resulting in altered confidence levels and increased prediction uncertainty \cite{11068155}. This aspect is particularly relevant in safety-critical and autonomous systems, where the predicted class alone is not the only decision variable, but is typically accompanied by confidence-related signals that influence downstream modules such as decision arbitration, sensor fusion, and fallback policies. In such architectures, even when the final decision remains unchanged, variations in the output distribution may lead to different system-level behaviours, including the activation of more conservative control strategies \cite{dong2025end}. From this perspective, the impact of fault injections cannot be fully captured by misclassification-based metrics alone. Instead, it is necessary to consider a more comprehensive set of fault-induced perturbations effect, since even non-switching faults can significantly reshape the prediction confidence \cite{8704548} and therefore affect the stability of subsequent decision-making stages under uncertainty induced by hardware variability. For this reason, we monitored also the variations in the output probability distribution produced by the DNN in the presence of faults. Specifically, we focused on the Top-1/Top-2 confidence margin ($R_\Delta$).

\subsection{Experimental results}

    For the evaluation, FI campaigns were set up using the following configurations: TS=32X32, NER per tile of 2\%, 4\%, 6\%, 8\%, and 10\%, and BER configurations of 20\%, 40\%, 60\%, 80\%, and 100\%. We used 10 SC configuration of Resnet50 (namely, CRBQ(1), CRBQ(2), CRBQ(3), CRBQ(6), CRBQ(9), CRBQ(12) and Entropic student with $\beta \in (0.64,1.28,2.56,5.12)$ from \cite{matsubara2023sc2}) as representative CNN models for the evaluation with the Imagenet dataset.
    All the experiments were conducted on a workstation HP Z2 G5 with a 20-core Intel i9-10800 CPU, 32 GB of RAM, and equipped with an RTX 3060TI NVIDIA’s Ampere GPU.

\subsubsection{Misclassification rate deviation}
\begin{figure*}[!t]
    \centering

    \begin{subfigure}{0.39\textwidth}
        \centering
        \includegraphics[width=\linewidth]{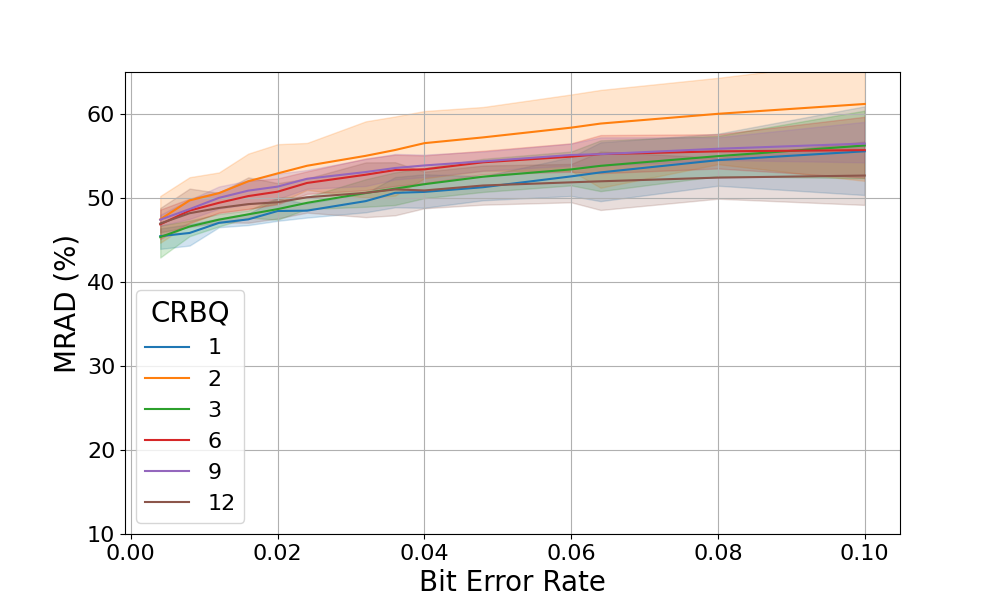}
        \caption{CRBQ configuration}
        \label{fig:mrad_crbq}
    \end{subfigure}
    \hspace{0.02\textwidth}
    \begin{subfigure}{0.39\textwidth}
        \centering
        \includegraphics[width=\linewidth]{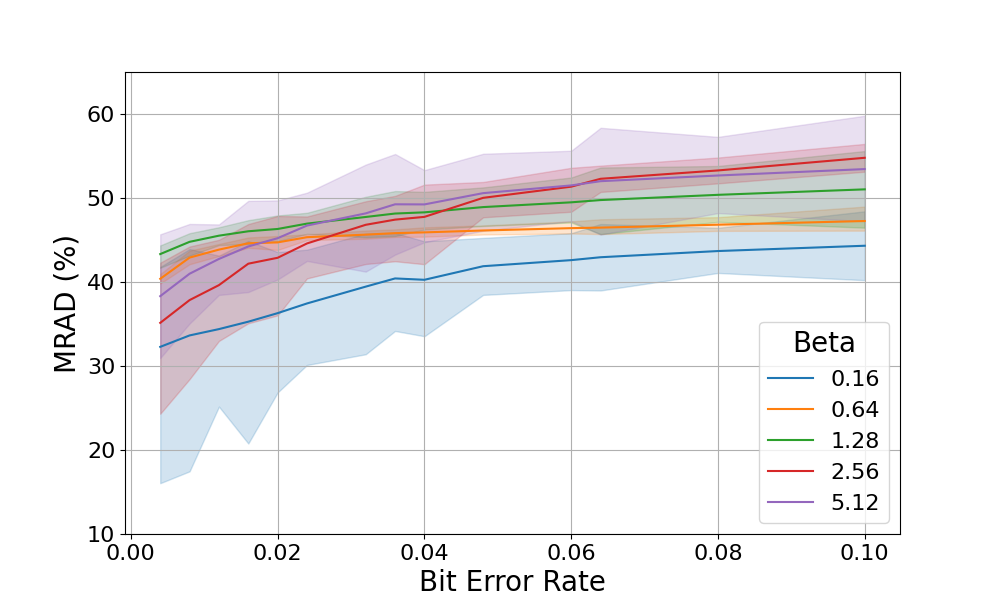}
        \caption{Entropic student configuration}
        \label{fig:mrad_entropic}
    \end{subfigure}
    \caption{MRAD as a function of the fault-corruption magnitude (BER).}
    \label{fig:mrad}
\end{figure*}

Figure~\ref{fig:mrad_crbq} reports the MRAD values measured for the analyzed CRBQ configurations as a function of the injected BER. Since MRAD quantifies the percentage of predictions that differ from the fault-free execution, higher values correspond to a larger number of fault-induced prediction changes and, consequently, a lower robustness against hardware faults. Across all configurations, MRAD increases with the BER, confirming that a higher fault rate progressively increases the probability of altering the SC DNN prediction. However, the growth rate strongly depends on the selected CRBQ configuration. CRBQ(2) consistently exhibits the highest MRAD values throughout the entire BER range, increasing from 47.45\% at BER~=~0.004 to 61.19\% at BER~=~0.1. Conversely, CRBQ(12) shows the lowest growth, ranging from 46.96\% to 52.67\%. The comparison among configurations reveals a clear separation at high BER values. At BER~=~0.1, CRBQ(12) achieves the lowest MRAD (52.67\%), followed by CRBQ(1) (55.56\%), CRBQ(6) (55.70\%), CRBQ(3) (56.25\%), CRBQ(9) (56.48\%), and CRBQ(2) (61.19\%). This behavior is due to the feature representations generated by CRBQ(12), preserving a higher degree of fault tolerance under bit-level perturbations. A possible explanation is that the corresponding compression configuration distributes the information content more uniformly across the encoded representation, reducing the impact of individual bit corruptions on the final prediction. An additional indicator of robustness is the sensitivity of MRAD to increasing BER. While CRBQ(2) exhibits a relative MRAD increase of approximately 28.9\% between the lowest and highest BER values, CRBQ(12) shows an increase of only 12.2\%. The remaining configurations lie between these two extremes. Therefore, considering both the absolute MRAD values and their evolution with BER, CRBQ(12) emerges as the most robust configuration among the analyzed alternatives.

Figure~\ref{fig:mrad_entropic} reports the MRAD obtained for the Entropic Student configurations as a function of the injected BER and the compression factor $\beta$. In this case, $\beta$ directly controls the compression applied to the intermediate feature map at the split point, with smaller values corresponding to more aggressive compression. The results show a clear dependence of MRAD on both BER and $\beta$. For all configurations, increasing the BER produces a monotonic increase in the percentage of altered predictions. At BER~=~0.004, MRAD ranges from 32.28\% for $\beta=0.16$ to 43.32\% for $\beta=1.28$, already producing a difference of more than 11 percentage points. At BER~=~0.1, the same gap increases further, with MRAD values ranging from 44.32\% ($\beta=0.16$) to 54.79\% ($\beta=2.56$). Interestingly, the most robust configuration corresponds to the strongest compression level ($\beta=0.16$). Across the entire BER range, this configuration consistently achieves the lowest MRAD values, increasing from 32.28\% to 44.32\%. In contrast, configurations characterized by larger $\beta$ values exhibit significantly higher MRAD values. For example, $\beta=2.56$ reaches 54.79\% at BER~=~0.1, while $\beta=5.12$ reaches 53.44\%. The difference between $\beta=0.16$ and $\beta=2.56$ therefore exceeds 10 percentage points at the highest BER level. The effect of the compression factor becomes even more evident when comparing configurations at a fixed BER. At BER~=~0.08, MRAD increases from 43.69\% for $\beta=0.16$ to 46.82\% for $\beta=0.64$, 50.38\% for $\beta=1.28$, 53.28\% for $\beta=2.56$, and 52.68\% for $\beta=5.12$. Most importantly, the variation induced by changing $\beta$ is larger than the variation induced by increasing BER. For instance, increasing BER from 0.004 to 0.1 causes an MRAD increase of approximately 12.0 percentage points for $\beta=0.16$, whereas changing the compression factor from $\beta=0.16$ to $\beta=2.56$ at BER~=~0.1 increases MRAD by approximately 10.5 percentage points. Similar trends are observed throughout the entire BER range. This indicates that the compression strategy adopted at the split point has a first-order effect on reliability, while BER acts mainly as a scaling factor that amplifies an already established sensitivity trend.

\subsubsection{Probability distribution deviation}

\begin{figure*}[!t]
    \centering

    \begin{subfigure}{0.39\textwidth}
        \centering
        \includegraphics[width=\linewidth]{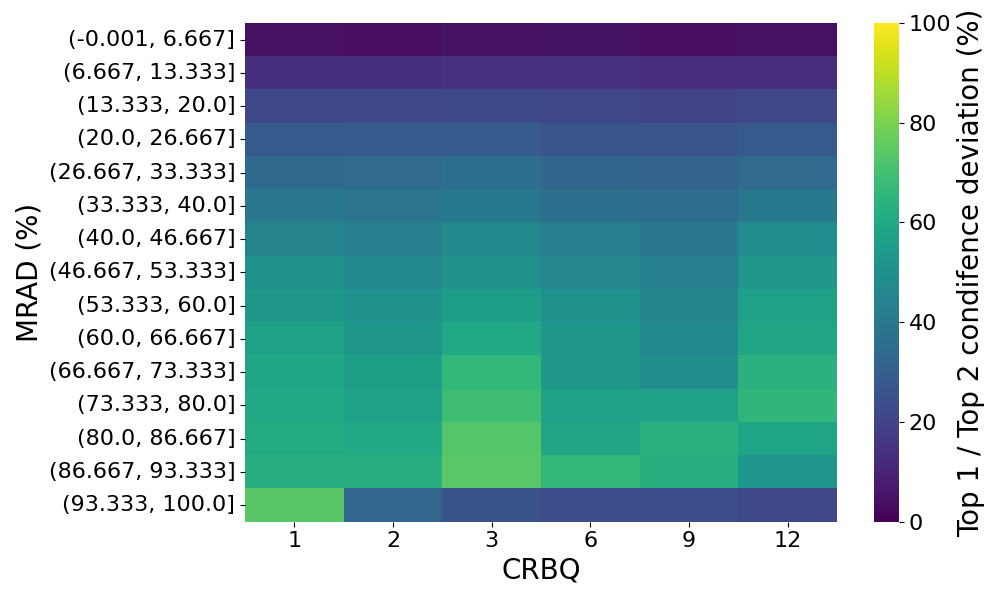}
        \caption{CRBQ configuration}
        \label{fig:gap_crbq}
    \end{subfigure}
    \hspace{0.02\textwidth}
    \begin{subfigure}{0.39\textwidth}
        \centering
        \includegraphics[width=\linewidth]{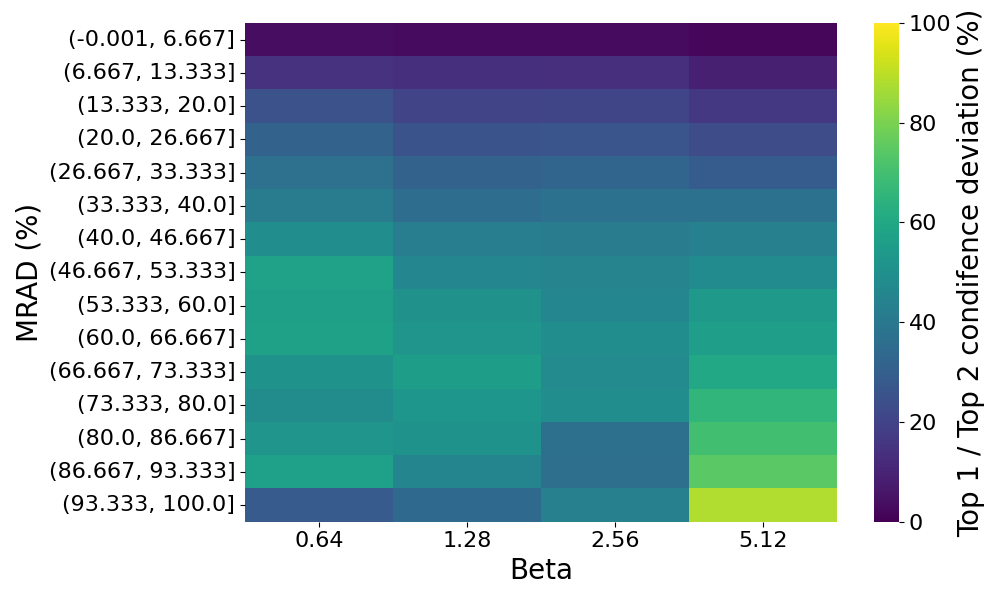}
        \caption{Entropic student configuration}
        \label{fig:gap_entropic}
    \end{subfigure}
    \caption{MRAD VS Magnitude of corruption (i.e., BER) VS Confidence Gap Deviation}
    \label{fig:reliability_confidence_gap}
\end{figure*}

Figure~\ref{fig:gap_crbq} shows $R_\Delta$ values with MRAD intervals for the analyzed CRBQ configurations. The heatmap allows correlating the frequency of prediction changes with the separation between the two most likely output classes normalized w.r.t. the golden $R_\Delta$. Across all configurations, the $R_\Delta$ generally increases with MRAD. For example, CRBQ(3) exhibits a margin of only 4.67 in the lowest MRAD interval ($0{-}6.67\%$), which progressively increases to 73.91 in the $86.67{-}93.33\%$ interval. Similar trends can be observed for CRBQ(1), whose margin increases from 4.06 to 63.19, and for CRBQ(12), whose margin increases from 4.28 to 65.40 before decreasing in the last MRAD interval. This behavior indicates a clear relationship between prediction instability and the separation of output confidence. As the percentage of corrupted predictions increases, the output probability distribution becomes more polarized, leading to larger differences between the most and second-most likely classes. A possible explanation is that hardware faults progressively distort the logits, amplifying the confidence assigned to a subset of classes while suppressing competing alternatives. Consequently, when a prediction changes, it is often associated with a strong shift in confidence rather than a near-boundary class exchange. Among the analyzed configurations, CRBQ(12) exhibits the most balanced behavior. Up to the MRAD interval $73.33{-}80.0\%$, its $R_\Delta$ remains comparable to or lower than those observed for CRBQ(1) and CRBQ(3), while also being associated with the lowest MRAD growth reported in the previous analysis. In contrast, CRBQ(3) reaches the largest margins, peaking at 73.91, whereas CRBQ(9) remains consistently below 64.03 throughout the entire range. The last MRAD interval ($93.33{-}100.0\%$) shows a margin collapse for all configurations except CRBQ(1), with CRBQ(12) dropping to 21.71 and CRBQ(3) to 25.65. This behavior suggests that once prediction corruption becomes dominant, the output distribution approaches a highly unstable regime in which confidence separation is no longer preserved. 

On the other side, figure~\ref{fig:gap_entropic} reports the average $R_\Delta$ for the Entropic Student configurations as a function of the MRAD interval and the compression factor $\beta$. A common trend across all compression factors is observed: larger MRAD values are generally associated with larger $R_\Delta$. For instance, for $\beta=0.64$, the margin increases from 3.45 in the lowest MRAD interval to 57.32 in the $60.0{-}66.67\%$ interval. Similarly, $\beta=5.12$ grows from 1.63 to 87.90, while $\beta=1.28$ increases from 3.02 to 55.38 before slightly decreasing at higher MRAD intervals. The dependence on the compression factor is also evident. Configurations with larger $\beta$ values tend to exhibit wider confidence margins at higher MRAD levels. For example, in the interval $86.67{-}93.33\%$, the $R_\Delta$ reaches 74.57 for $\beta=5.12$, compared with 56.46 for $\beta=0.64$, 45.49 for $\beta=1.28$, and 36.33 for $\beta=2.56$. In the highest MRAD interval ($93.33{-}100.0\%$), $\beta=5.12$ reaches the largest observed margin of 87.90, while the remaining configurations remain below 44. These results indicate that increasing MRAD is accompanied by a progressive widening of the confidence gap between the most probable classes. This is because faults do not simply induce random class exchanges. Instead, they progressively reshape the output probability distribution, causing the DNN to assign increasingly dominant confidence values to the selected class while suppressing alternative predictions. As a consequence, prediction changes are often associated with highly polarized output distributions. The influence of the compression factor remains consistent with the MRAD analysis discussed previously. Configurations with more aggressive feature-map compression generally exhibit lower MRAD values and slower growth in confidence margin. Conversely, larger $\beta$ values tend to produce both higher MRAD and larger $R_\Delta$. Indeed, more compressed feature representations preserve a larger number of fault-sensitive activation patterns, allowing hardware-induced perturbations to propagate more effectively toward the output layer and amplify the resulting confidence redistribution. As the percentage of altered predictions increases (i.e., MRAD), the confidence gap between the most likely classes widens. Therefore, the increase in MRAD is not only associated with a higher number of prediction changes but also with increasingly pronounced modifications to the output confidence distribution.

\section{Security}
\label{sec:security}
This section describes the strategy for evaluating the security of SC architectures in Edge computing.
Moreover, we report and analyze the experimental results.
\subsection{Evaluation strategy}
\label{subsec:sec_eval_strat}

Hardware security is a central challenge for the distributed devices that run Split Computing inference. Indeed, in the SC deployment scenario the Head model executes on a resource-constrained mobile device that is physically exposed and does not offer the protection guarantees of a trusted edge or cloud server \cite{xu2023edge}. Because the Head model is deployed on this exposed device, it can become available to an attacker who exploits the device's hardware security vulnerabilities. With access to the Head and to the intermediate feature maps it produces, the attacker can craft the attack without altering the model parameters.

Adversarial attacks simulations are one way to reproduce this behavior. The attacker introduces a small, carefully crafted perturbation into the input so that the model produces an incorrect prediction, while the Head and Tail weights remain fixed \cite{goodfellow2015explaining, szegedy2014intriguing, akhtar2018threat}. We reflect this threat by computing the gradient of an intermediate Head feature map and using it to build a noise filter that is applied to the input image, resorting to the threat model presented in \cite{esposito2025aaclassification, esposito2026ai}. The transmitted representation is never corrupted directly, so the evaluation stays on the device-side attack surface and remains distinct from threats that exploit the wireless link between the Head and the Tail.

The attack follows the gradient-based adversarial model introduced in \cite{esposito2025aaclassification, esposito2026ai}, which assumes a gray-box adversary that can read the Head model but has no access to the Tail. The adversary crafts the perturbation from the bottleneck representation produced at the end of the Head, i.e., the feature map that the device transmits to the Tail. Because this representation carries all the information that reaches the Tail, suppressing its magnitude removes the evidence needed for a correct decision and drives the model toward a wrong prediction. This setting justifies the adversarial input attack as a faithful proxy for an attacker that poisons the data consumed by the exposed Head device.

The attack is iterative. Starting from the clean image, it computes the gradient of the bottleneck representation with respect to the input and updates the input by a fixed amount $\epsilon$ in the direction given by the sign of that gradient, keeping every pixel within the valid image range. The step size $\epsilon$ is the perturbation budget and sets the strength of the attack, playing the same role that the Bit Error Rate plays in the fault campaign. The procedure runs for at most 33 iterations and stops as soon as the predicted class changes with respect to the golden inference. We sweep $\epsilon \in \{0.001, 0.003, 0.005, 0.01\}$, so that the attack effect can be read as a function of the perturbation magnitude.

We assess security with two metrics, both already well-established in the literature. A security threat reaches the model output in two ways. The first is a change of the predicted class, which is the critical outcome because it corrupts the decision the system acts on. We capture it with the Attack Success Rate (ASR) \cite{carlini2017towards}, the fraction of images whose attacked top-1 prediction differs from the golden one, counted over all 625 images so that samples already misclassified by the clean model are not excluded. The second effect is more subtle but still security-relevant: even when the predicted class stays the same, the attack can push the prediction toward the decision boundary by narrowing the separation between the two most likely classes, which leaves the model one small perturbation away from a wrong decision. We capture this effect with the Top-1/Top-2 confidence gap (or $S\_\Delta$) \cite{11068155}, the distance between the two most probable classes, measured only on the images whose prediction is unchanged and reported as a relative reduction with respect to the golden gap. Restricting it to unchanged predictions keeps the comparison on the same class and isolates this confidence effect from the prediction change already counted by ASR.

We aligned these two metrics with the ones used on the reliability side, the accuracy degradation and the same Top-1/Top-2 confidence gap, so that the results of the two campaigns are comparable and can be combined into the joint assessment.\footnote{The implementation used for the security evaluation is available in this repository: \url{https://github.com/Reliable-and-Secure-Systems/DFTS_JASPER_Security}}

\subsection{Experimental results}

\subsubsection{Attack success across the perturbation budget}

\begin{figure*}[!t]
    \centering
    \begin{subfigure}{0.39\textwidth}
        \centering
        \includegraphics[width=\linewidth]{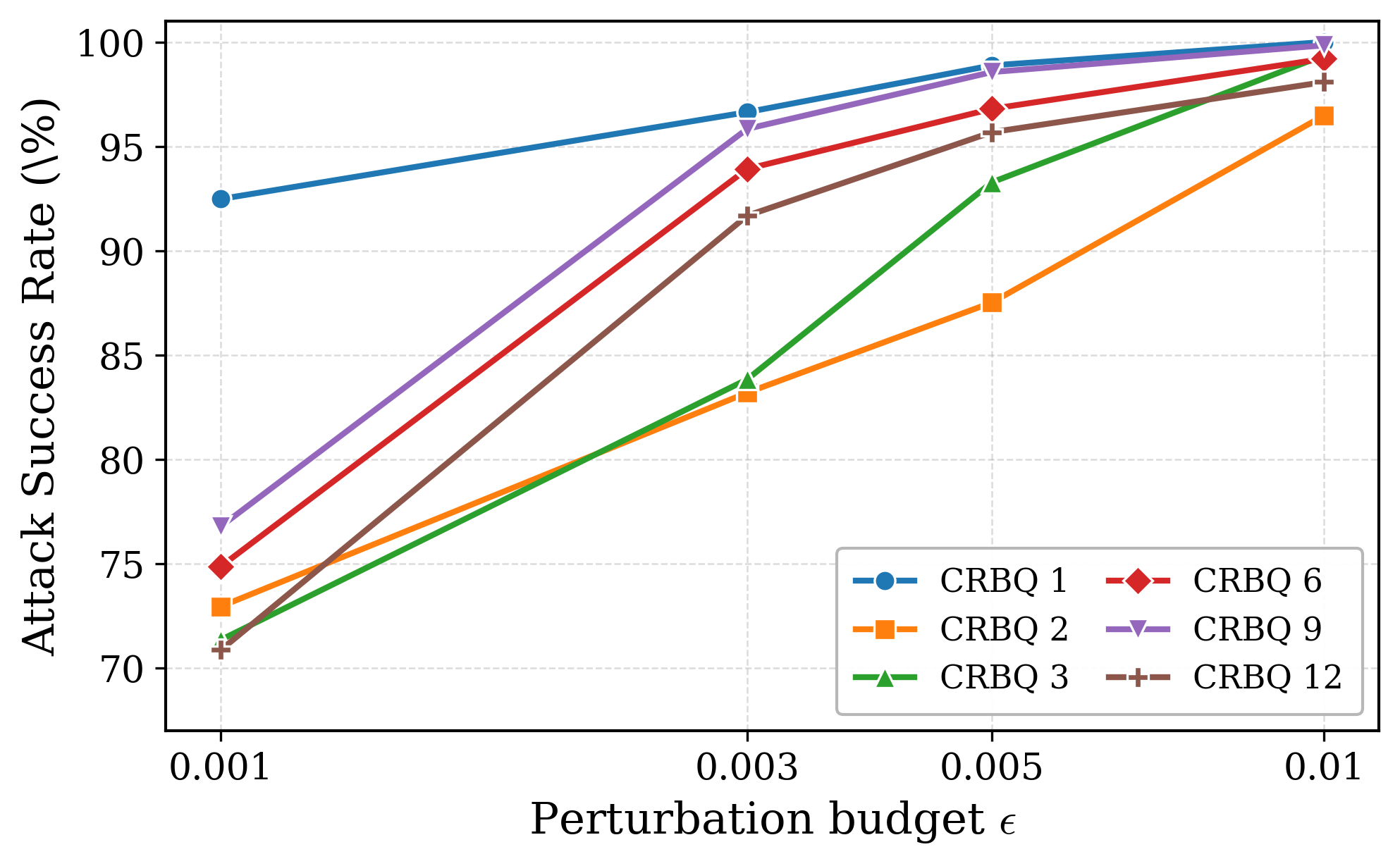}
        \caption{ASR of the input attack for the CRBQ configurations as a function of the perturbation budget $\epsilon$, expressed as a percentage. Each curve corresponds to one configuration.}
        \label{fig:security_asr_vs_epsilon_crbq}
    \end{subfigure}
    \hspace{0.02\textwidth}
    \begin{subfigure}{0.39\textwidth}
        \centering
        \includegraphics[width=\linewidth]{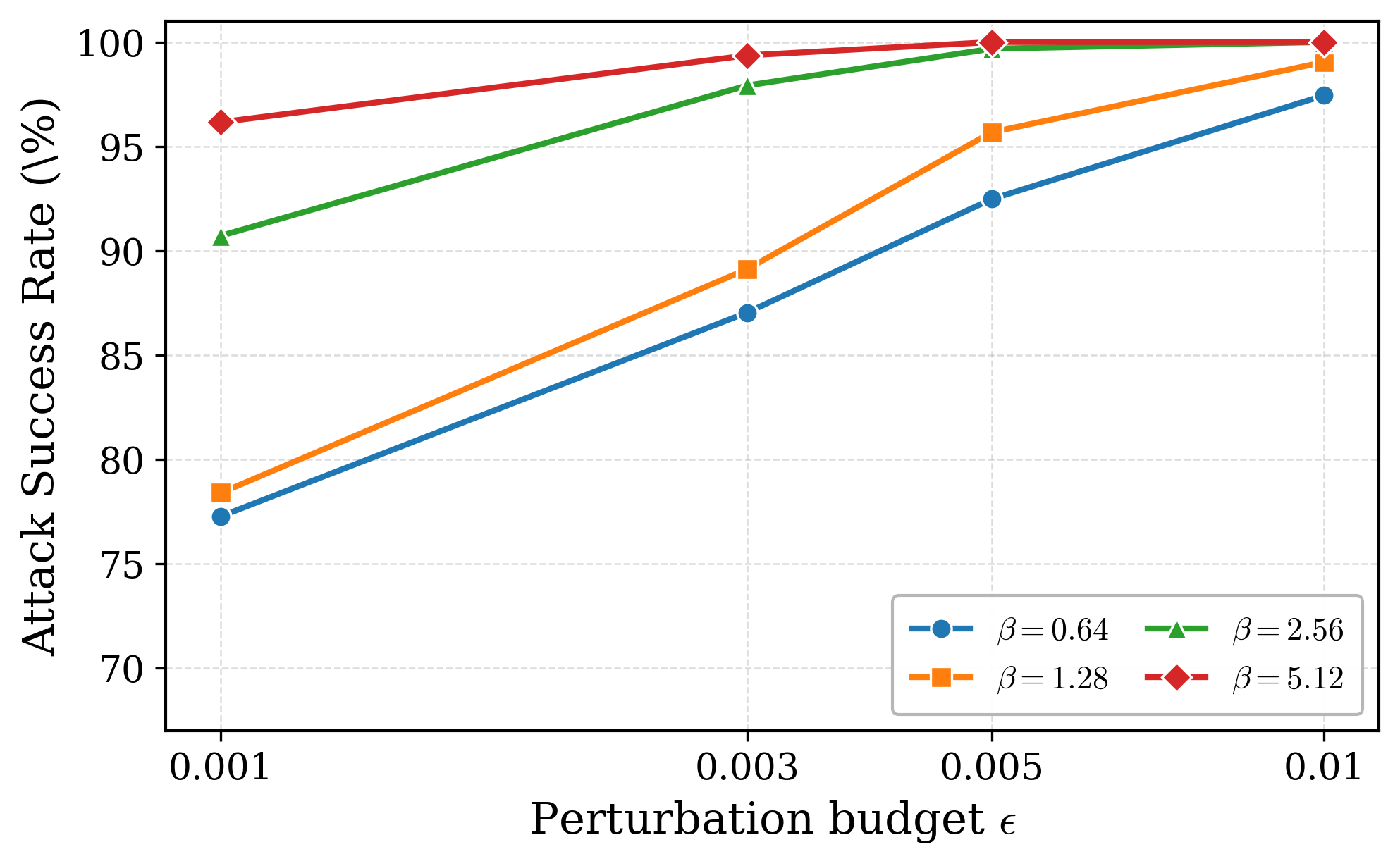}
        \caption{ASR of the input attack for the Entropic Student configurations as a function of $\epsilon$, expressed as a percentage. Each curve corresponds to one value of $\beta$.}
        \label{fig:security_asr_vs_epsilon_entropic}
    \end{subfigure}
    \caption{ASR VS magnitute of corruption (i.e., $\epsilon$)}
    \label{fig:asr}
\end{figure*}

Figure~\ref{fig:security_asr_vs_epsilon_crbq} reports the prediction-change ASR of the CRBQ configurations as a function of the perturbation budget, with the configuration in the legend and the metric expressed as a percentage. ASR grows monotonically with $\epsilon$ for every configuration. Averaged over the six CRBQ models, it rises from 76.6\% at $\epsilon=0.001$ to 90.9\% at $\epsilon=0.003$, 95.1\% at $\epsilon=0.005$, and 98.8\% at $\epsilon=0.01$, while the corresponding mean attacked accuracy falls from 23.8\% to 10.4\%, 6.4\%, and 3.1\%, against a mean golden accuracy of 83.6\%. At the reference budget $\epsilon=0.003$, the per-configuration ASR is 96.6\%, 83.2\%, 83.8\%, 93.9\%, 95.8\%, and 91.7\% for CRBQ 1, 2, 3, 6, 9, and 12, respectively. CRBQ 2 and CRBQ 3 remain the most resistant at every budget, keeping attacked accuracy above 10\% at $\epsilon=0.003$, whereas CRBQ 1, 6, 9, and 12 already fall below it. CRBQ 1 is the most sensitive of all: its golden accuracy is the lowest of the set, 64.6\% against 85.0\% to 88.8\% for the others, and its ASR reaches 100\% at $\epsilon=0.01$. This separation shows that the compression configuration, and not only the perturbation budget, governs how easily the model is fooled.

The Entropic Student configurations follow the same trend in Fig.~\ref{fig:security_asr_vs_epsilon_entropic}. ASR grows with $\epsilon$ for every value of $\beta$ and saturates near 100\% at the largest budget. For $\beta=0.64$ it rises from 77.3\% at $\epsilon=0.001$ to 87.0\%, 92.5\%, and 97.4\%, and for $\beta=1.28$ from 78.4\% to 89.1\%, 95.7\%, and 99.0\%. The two most compressed models are already close to saturation at small budgets: $\beta=2.56$ moves from 90.7\% to 100\% and $\beta=5.12$ from 96.2\% to 100\%. The ordering is consistent across the sweep, with stronger compression (higher $\beta$) giving higher ASR. Because the golden accuracies differ across $\beta$, from 86.9\% at $\beta=0.64$ down to 67.2\% at $\beta=5.12$, we compare the configurations through ASR rather than absolute attacked accuracy.

\subsubsection{Output confidence and uncertainty}

\begin{figure*}[!t]
    \centering

    \begin{subfigure}{0.39\textwidth}
        \centering
        \includegraphics[width=\linewidth]{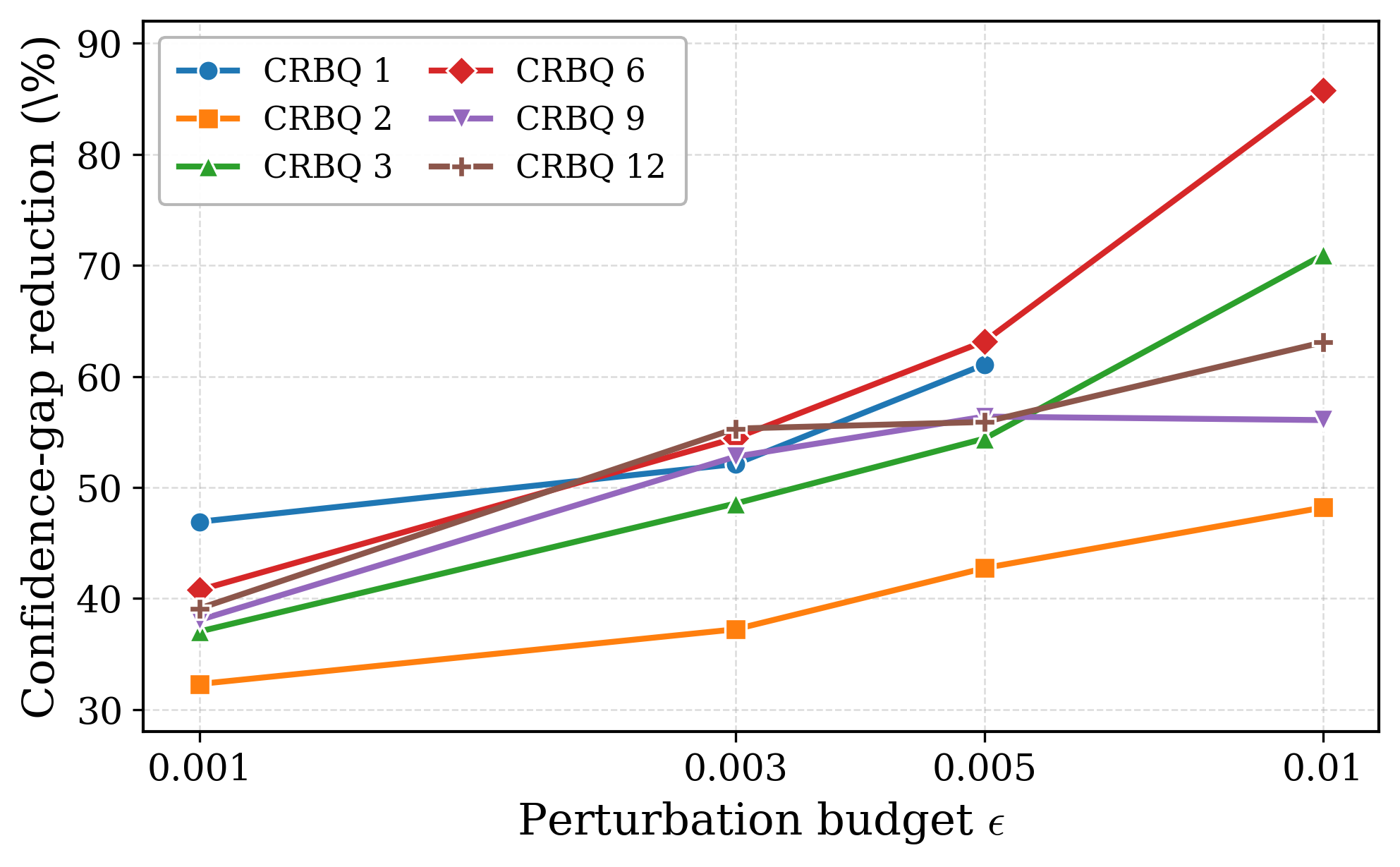}
        \caption{Relative $S_\Delta$ reduction for the CRBQ configurations, measured on the images whose prediction is unchanged, as a function of $\epsilon$.}
        \label{fig:security_gap_vs_epsilon_crbq}
    \end{subfigure}
    \hspace{0.02\textwidth}
    \begin{subfigure}{0.39\textwidth}
        \centering
        \includegraphics[width=\linewidth]{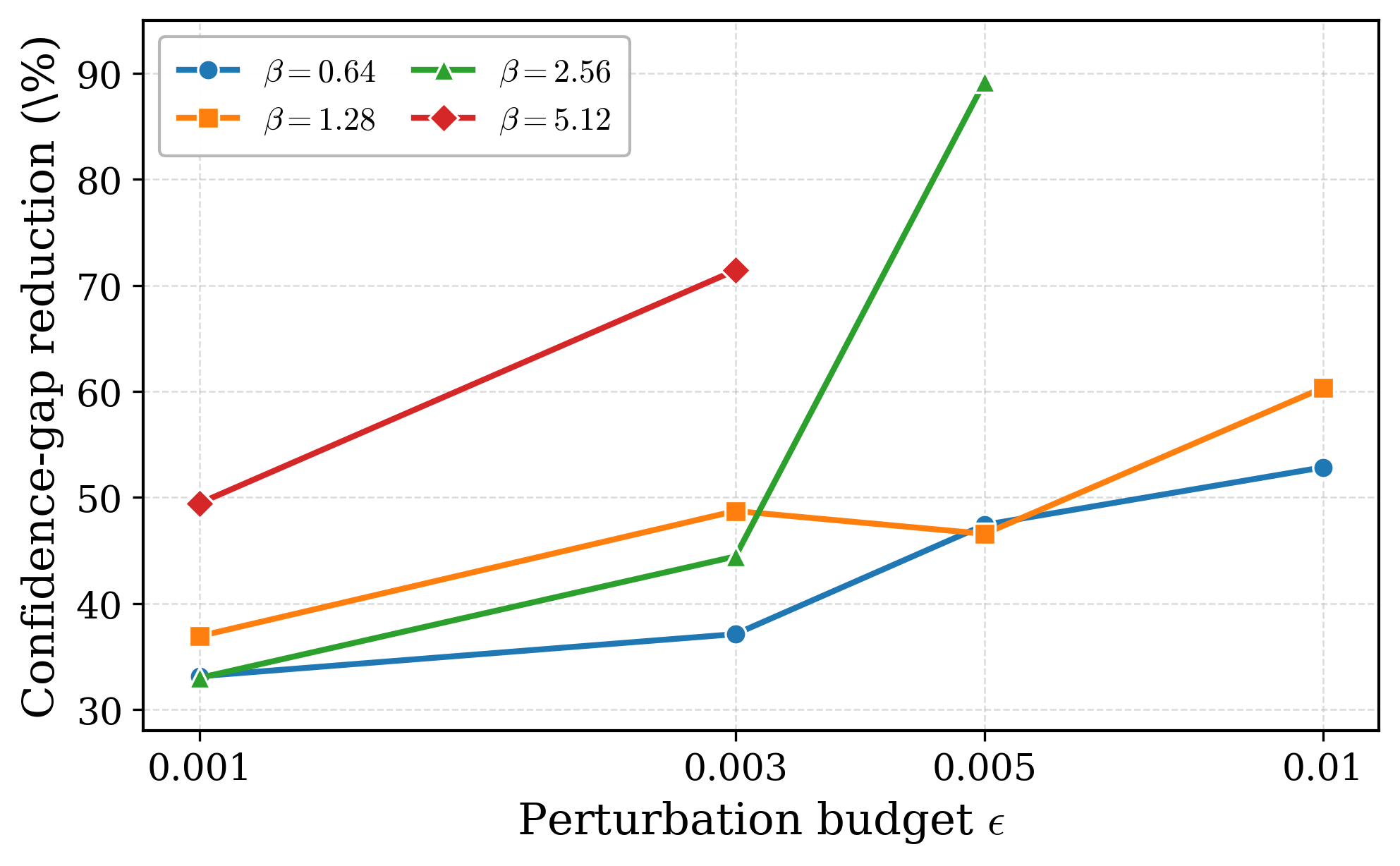}
        \caption{Relative $S_\Delta$ reduction for the Entropic Student configurations, measured on the images whose prediction is unchanged, as a function of $\epsilon$.}
        \label{fig:security_gap_vs_epsilon_entropic}
    \end{subfigure}
    \caption{Normalized $S_\Delta$ VS magnitute of corruption (i.e., $\epsilon$)}
    \label{fig:security_confidence_gap}
\end{figure*}

Figure~\ref{fig:security_gap_vs_epsilon_crbq} reports the $S_\Delta$ reduction on the unchanged-prediction subset as a function of $\epsilon$, with the configuration in the legend. The reduction grows with the budget. Averaged over the six CRBQ configurations it is 39.0\% at $\epsilon=0.001$, 50.1\% at $\epsilon=0.003$, and 55.6\% at $\epsilon=0.005$. At the reference budget $\epsilon=0.003$ the per-configuration values run from 37.2\% for the most resistant model, CRBQ 2, to 55.3\% for CRBQ 12. The unchanged subset shrinks quickly as more predictions flip, from between 47 and 182 images per configuration at $\epsilon=0.001$ to single digits at $\epsilon=0.01$, so the largest-budget points are based on too few samples to be reliable for the most sensitive models. The predictions that survive still lose a large part of their confidence gap, which shows that the attack erodes decision confidence even when it does not change the class. This is the security counterpart of the Safe-SDC behavior observed under faults.

The Entropic Student shows the same behavior in Fig.~\ref{fig:security_gap_vs_epsilon_entropic}. On the unchanged subset the confidence-gap reduction grows with $\epsilon$, from 33.1\% to 47.4\% for $\beta=0.64$ and from 36.9\% to 48.7\% for $\beta=1.28$ between $\epsilon=0.001$ and $\epsilon=0.005$. The two most compressed models leave almost no unchanged predictions once the budget grows, so the curves for $\beta=2.56$ and $\beta=5.12$ are defined only at the smaller budgets. The trend matches the CRBQ case: the attack lowers the decision confidence of the predictions it does not flip.

Taken together, the security results show that the compression strategy changes both the probability of flipping the prediction and the confidence left on the samples that survive the attack. For the CRBQ family, CRBQ 2 and CRBQ 3 provide the strongest resistance in the tested sweep, with lower ASR and smaller confidence-gap reduction than the other CRBQ configurations. For the Entropic Student family, higher $\beta$ values are more vulnerable to this attack: $\beta=5.12$ reaches saturation earlier and leaves too few unchanged predictions at larger budgets for a stable confidence-gap estimate. Therefore, compression cannot be treated as uniformly beneficial from the security perspective. Its effect depends on the compression mechanism and must be evaluated together with the perturbation budget.

\section{Joint Reliability and Security Assessment}
\label{sec:joint_metric}

As described in Section \ref{sec:introduction}, the main aim of this work is to consider reliability and security as a single feature of Robustness that can better explain the vulnerability of the system under evaluation and the impact of possible threats.
The work is mainly motivated by the fact that improving security does not necessarily mean improving the reliability of the system, improving reliability may negatively affect the security feature of the system \cite{7574542}.

\begin{figure*}[!t]
    \centering
    \begin{subfigure}{0.450\textwidth}
        \centering
        \includegraphics[width=\linewidth]{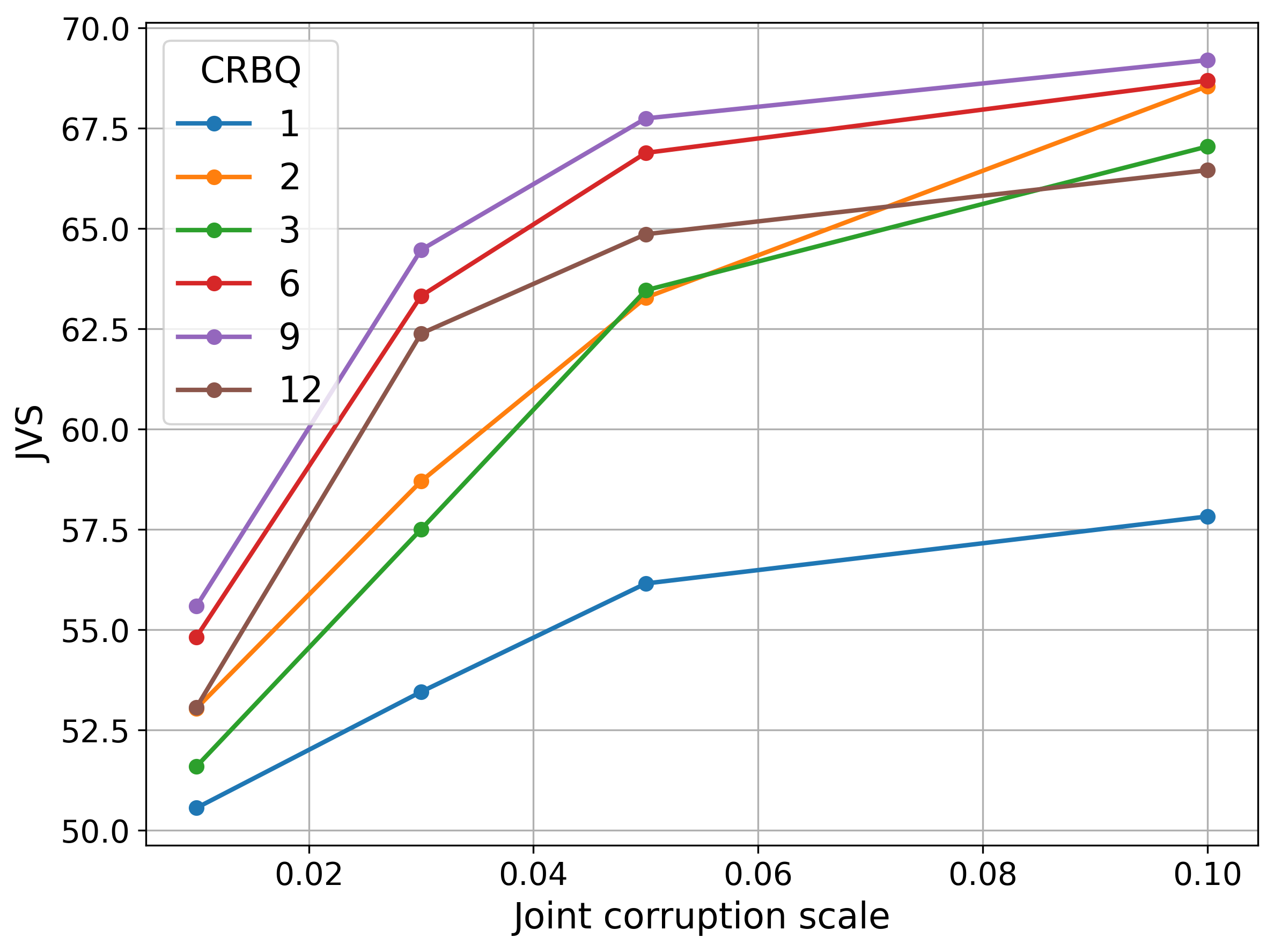}
        \caption{CRBQ configuration.}
        \label{fig:jvs_crbq}
    \end{subfigure}
    \hspace{0.02\textwidth}
    \begin{subfigure}{0.450\textwidth}
        \centering
        \includegraphics[width=\linewidth]{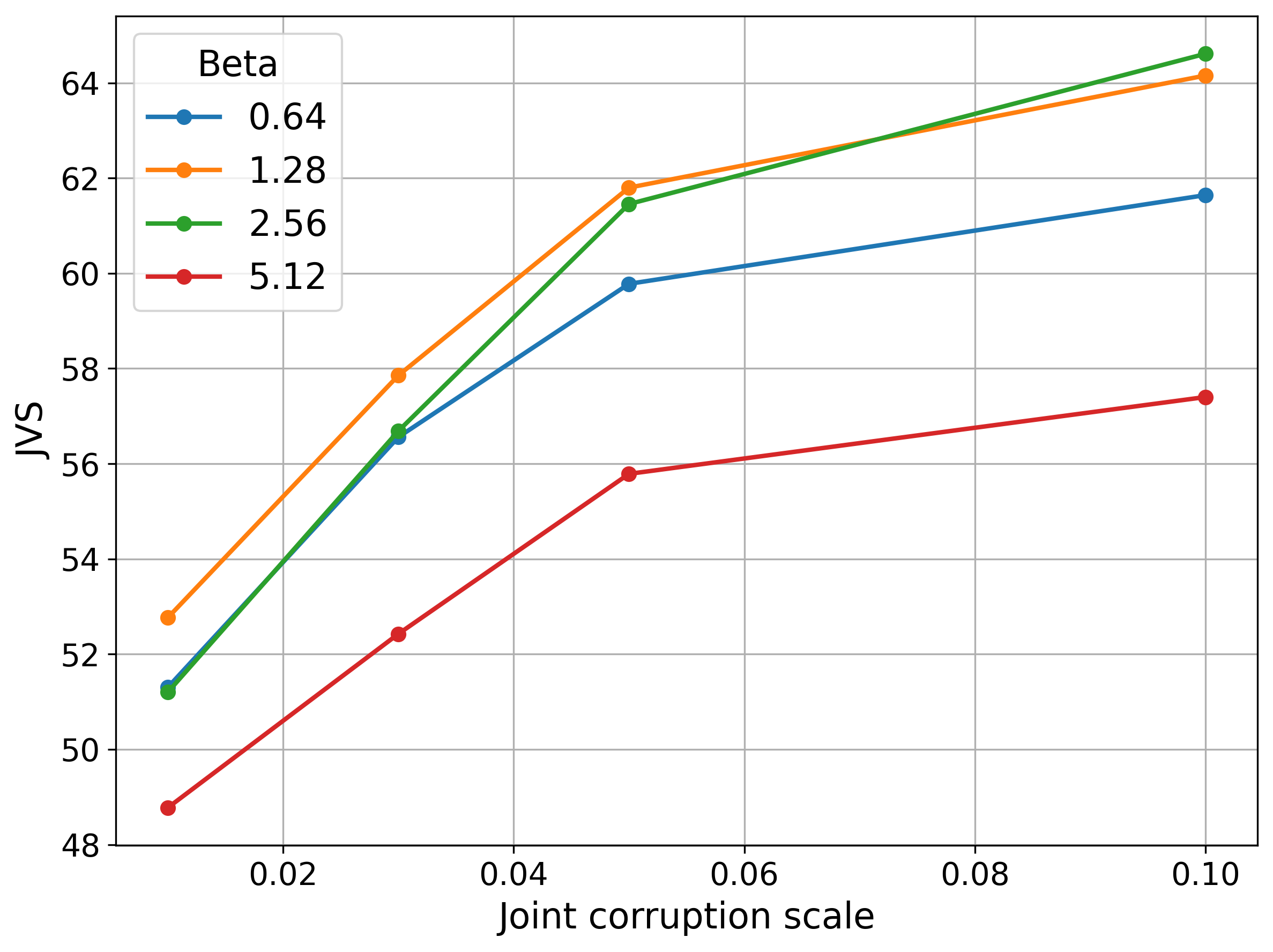}
        \caption{Entropic student configuration.}
        \label{fig:jvs_entropic}
    \end{subfigure}
    \caption{JVS for different perturbation patterns under different security configurations.}
    \label{fig:jvs}
\end{figure*}

\begin{figure*}[!t]
    \centering
    \begin{subfigure}{0.45\textwidth}
        \centering
        \includegraphics[width=\linewidth]{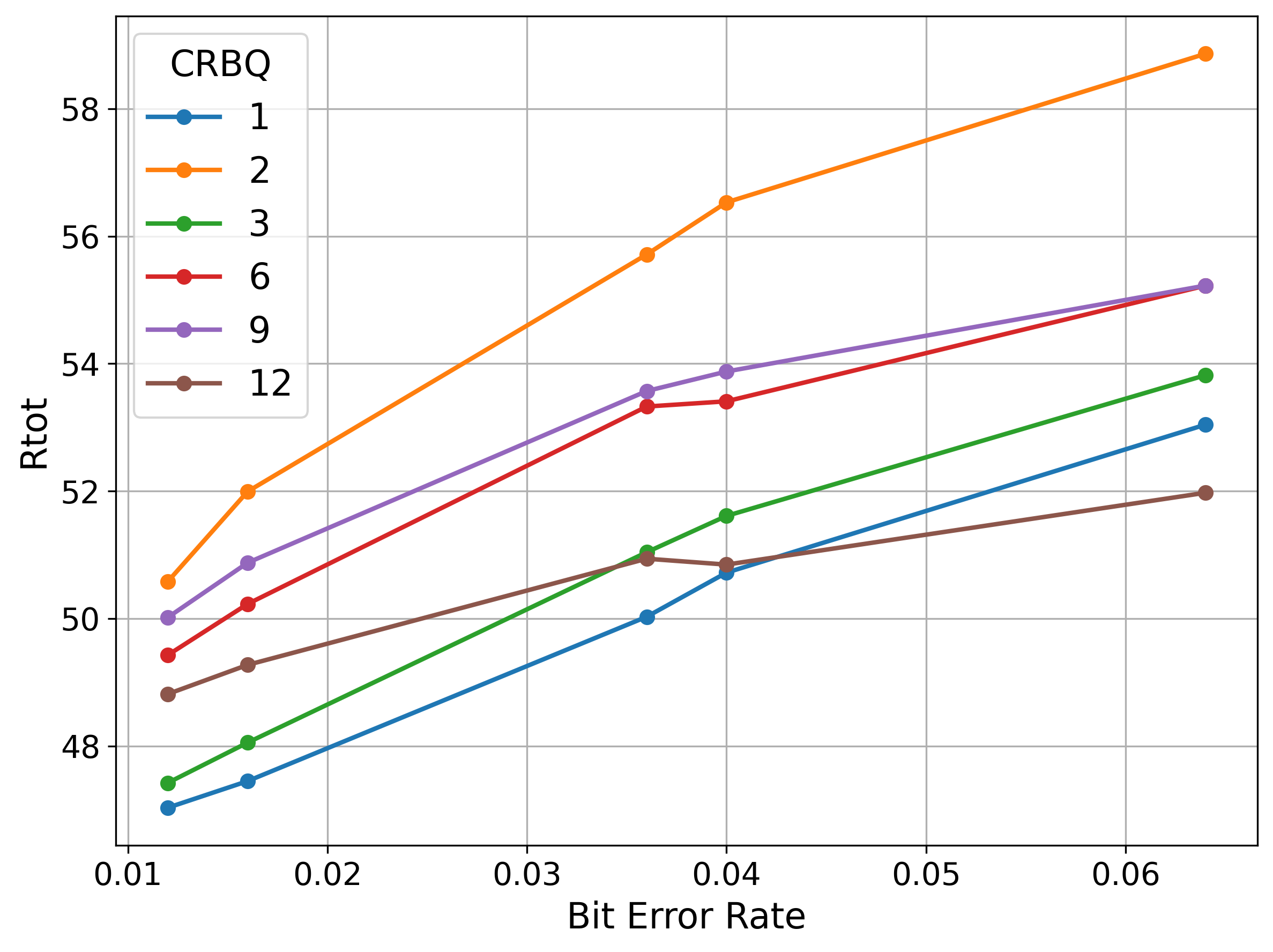}
        \caption{$R_{tot}$, CRBQ.}
        \label{fig:Rtot_crbq}
    \end{subfigure}
    \hspace{0.025\textwidth}
    \begin{subfigure}{0.45\textwidth}
        \centering
        \includegraphics[width=\linewidth]{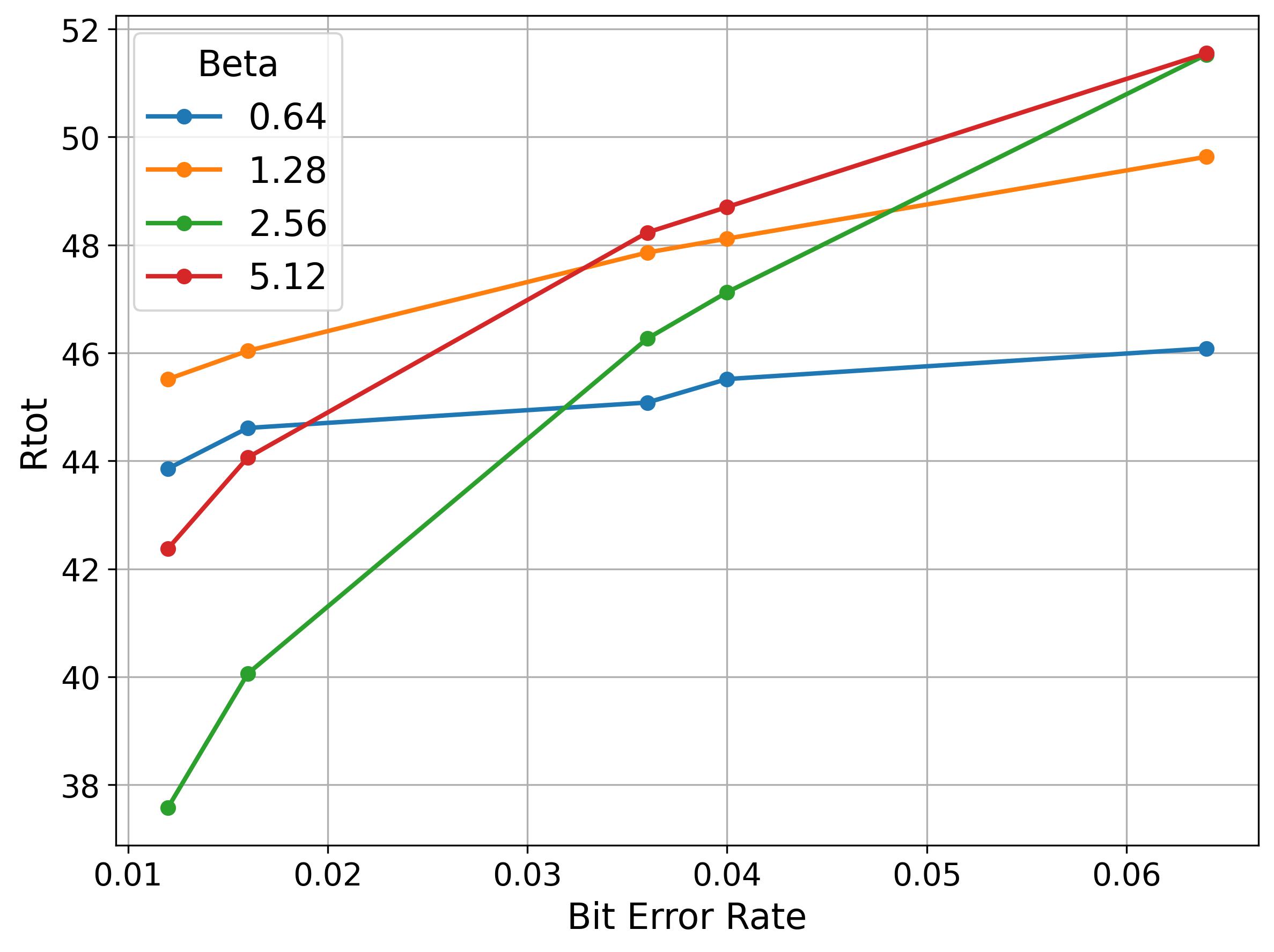}
        \caption{$R_{tot}$, Entropic Student.}
        \label{fig:Rtot_entropic}
    \end{subfigure}
    \par\smallskip
    \begin{subfigure}{0.45\textwidth}
        \centering
        \includegraphics[width=\linewidth]{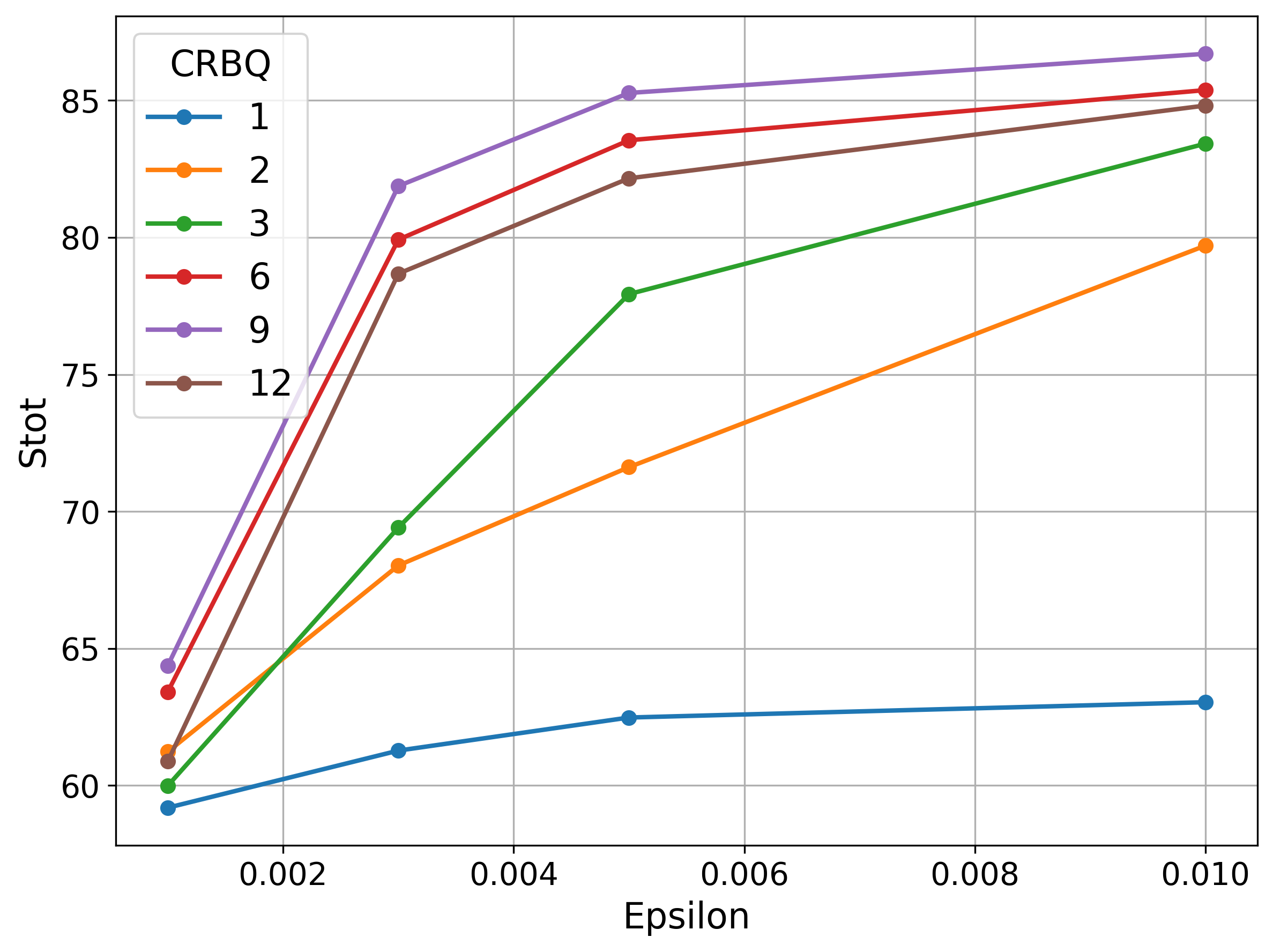}
        \caption{$S_{tot}$, CRBQ.}
        \label{fig:Stot_crbq}
    \end{subfigure}
    \hspace{0.025\textwidth}
    \begin{subfigure}{0.45\textwidth}
        \centering
        \includegraphics[width=\linewidth]{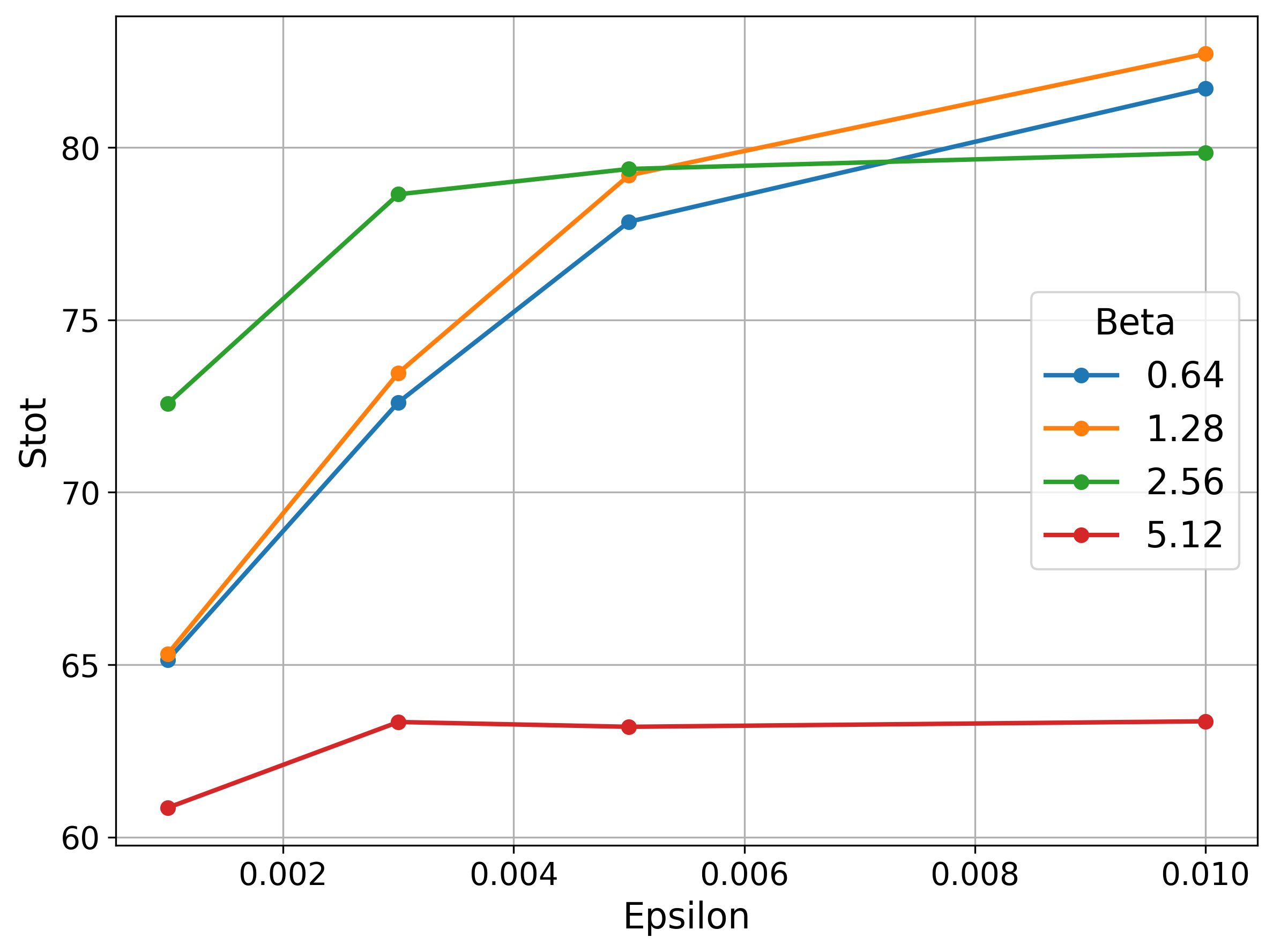}
        \caption{$S_{tot}$, Entropic Student.}
        \label{fig:Stot_entropic}
    \end{subfigure}
    \caption{Evolution of reliability ($R_{tot}$) and security ($S_{tot}$) vulnerability metrics under different corruption levels and Split Computing configurations. The first two panels report reliability degradation as a function of Bit Error Rate (BER), while the last two report security degradation as a function of adversarial perturbation magnitude ($\epsilon$).}
    \label{fig:Rtot_Stot}
\end{figure*}

\begin{figure*}[!t]
    \centering
    \begin{subfigure}{0.45\textwidth}
        \centering
        \includegraphics[width=\linewidth]{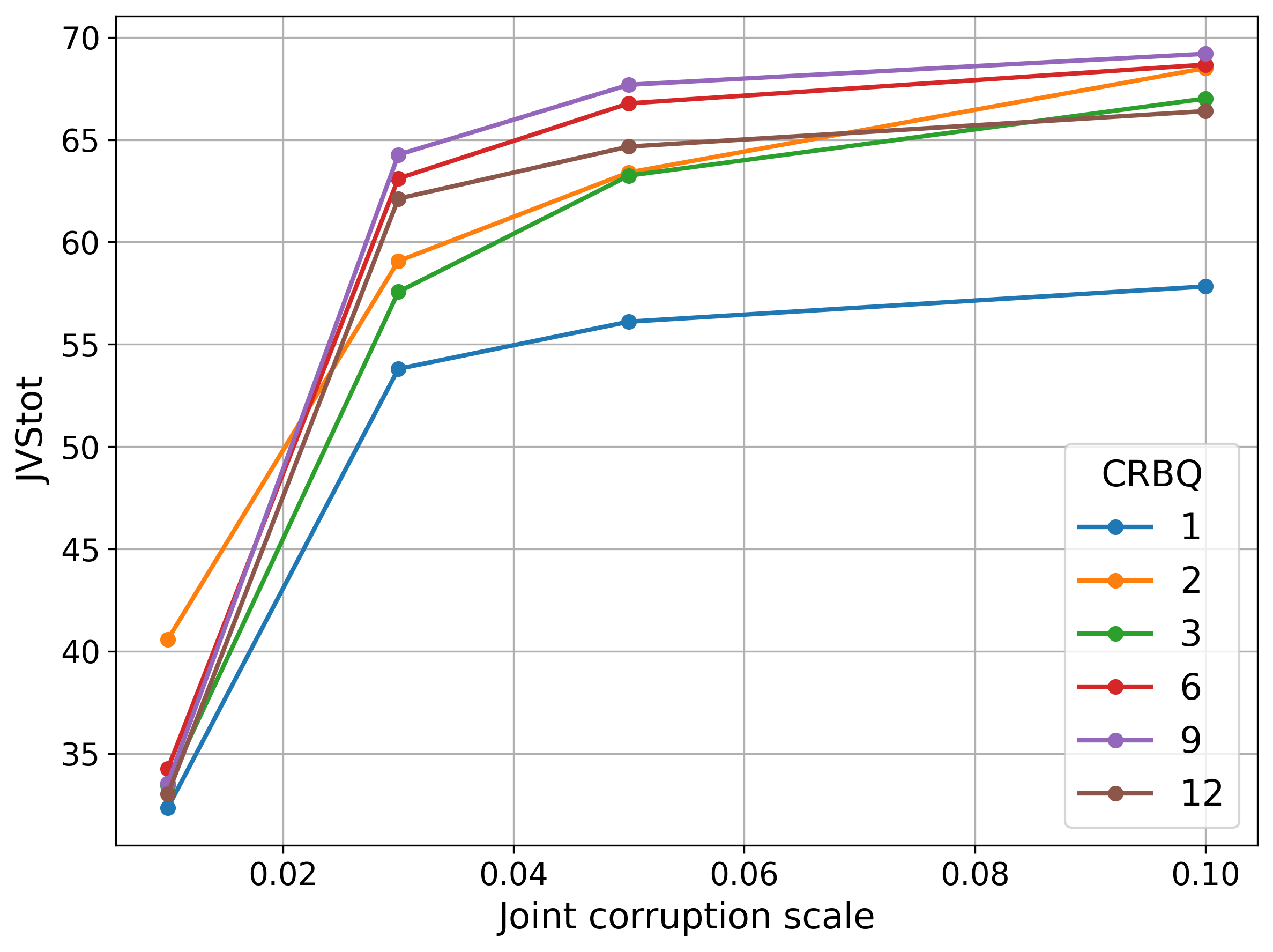}
        \caption{CRBQ configuration.}
        \label{fig:JVStot_crbq}
    \end{subfigure}
    \hspace{0.02\textwidth}
    \begin{subfigure}{0.45\textwidth}
        \centering
        \includegraphics[width=\linewidth]{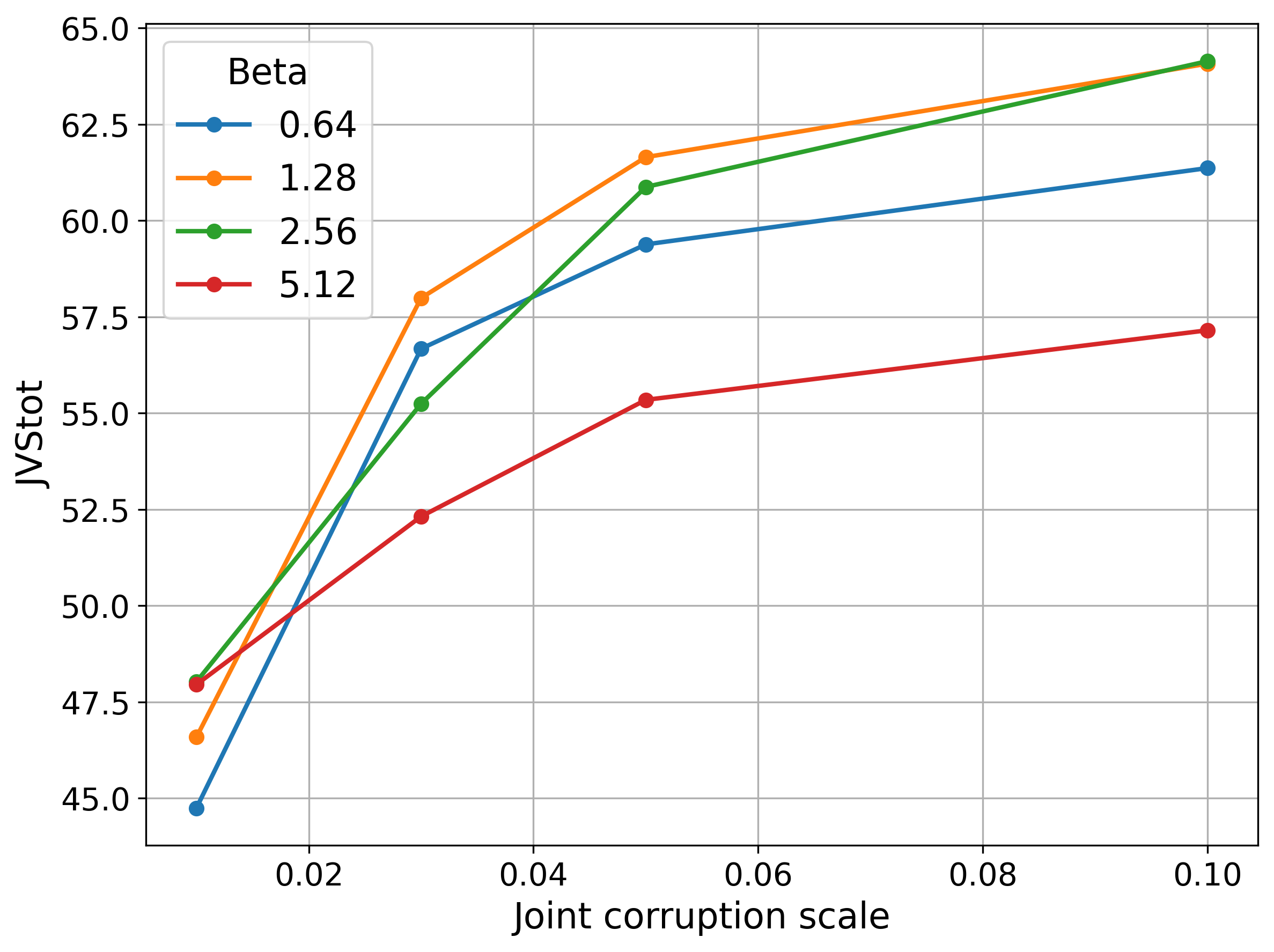}
        \caption{Entropic Student configuration.}
        \label{fig:JVStot_entropic}
    \end{subfigure}
    \caption{Extended JVS joint reliability and security metric for different configurations and target layers. It accounts for both accuracy degradation and confidence score degradation.}
    \label{fig:JVStot}
\end{figure*}

\subsection{Evaluating Reliability and Security as a joint SC system robustness feature}

The first proposed Joint metric, the Joint Vulnerability Score Metric (JVS), is defined to capture the trade-off between model vulnerability to faults and to adversarial attacks. It combines the reliability term MRAD, derived from the accuracy degradation after FI campaigns, as described in the section \ref{sec:reliability}, and the security term ASR, derived from the adversarial attack simulations. 

The original reliability and security vulnerability metrics fall within the following range: $x \in [0, 100]$.

Both terms represent the vulnerability score of the SC model, which is a measure of model misclassification. This makes them directly comparable and suitable for joint aggregation. However, the model misclassification is due to different causes.

The Joint metric is computed as follows:

\begin{equation}
\mathrm{JVS} = \sqrt{MRAD \cdot ASR}
\label{eq:JVS}
\end{equation}

Where MRAD \(\in [0,100] \) represents the normalized reliability vulnerability score of the model under fault conditions and ASR \(\in [0,100] \) quantifies the system's vulnerability under adversarial attack simulations. A high JVS indicates high model vulnerability, either to injected faults or adversarial attack simulations.

In practice, MRAD and ASR are independently collected and subsequently unified into a single dataset before computing the proposed metric. Since the FI campaigns is parameterized through the Bit Error Rate (BER), while the adversarial attack simulation is parameterized through the adversarial perturbation magnitude $\epsilon$, both domains are first projected onto the common JVS. 

Reliability and security measurements and JVS value are aggregated by averaging the reliability indicators. Finally, the two datasets are merged using the pair $(\mathrm{JVS}, \mathrm{Configuration})$ as common keys, producing a unified dataset from which all the proposed joint metrics are computed. This procedure enables a direct comparison and aggregation of reliability and security results despite their originally different experimental parameterizations.

The proposed JVS metric considers reliability and security together to determine whether the prediction changes under fault injection or an adversarial attack. However, it does not account for confidence variation associated with predictions under threat experiments. Two configurations may have the same JVS value but exhibit significantly different levels of confidence variation under threat. This detail is relevant because a model configuration with a small difference between the top-1 and top-2 class probabilities is more likely to produce misclassifications.

Starting from the original reliability and security vulnerability metrics, the confidence-aware reliability and security terms can then be obtained by combining the outcome-based and confidence-based information through a weighted average:

\begin{equation}
R_{tot}=\alpha \cdot MRAD + (1-\alpha)\cdot(100 - R_{\Delta})
\label{eq:Rtot}
\end{equation}

\begin{equation}
S_{tot}=\beta \cdot ASR + (1-\beta)\cdot(100 - S_{\Delta})
\label{eq:Stot}
\end{equation}

where $\alpha$ and $\beta$ are weighting coefficients that regulate the contribution of prediction correctness and prediction confidence under the idea to track in both contribution of the threat, the misclassifications, and the variation of the confidence level, based on their corresponding contribution to the behavior of the SC DNN under threats.
The complementary terms $(100-R_{\Delta})$ and $(100-S_{\Delta})$ are used to convert prediction confidence into a measure of uncertainty. Since $R_{\Delta}$ (as described in Section \ref{subsec:rel_eval_strat}) and $S_{\Delta}$ (as described in Section \ref{subsec:sec_eval_strat}) represent confidence percentages associated with the Top-1 prediction. By considering their complements, the proposed formulation instead quantifies the percentage of confidence lost, how much the model is uncertain about the correctness of its prediction. This choice ensures that increasing values of $R_{tot}$ and $S_{tot}$ consistently reflect increasing levels of reliability and security degradation, respectively.

In the case of the Fault injection, $\alpha$ is the rate of prediction changes w.r.t. the golden prediction over the total number of performed predictions.
Regarding the Adversarial Attack, the $\beta$ is the ASR.
Both $\alpha$ and $\beta$ are in the range \( \in [0,1] \)

Finally, the extended JVS metric is derived from the JVS formulation (see equation \ref{eq:JVS}). The terms MRAD and ASR are substituted with the more comprehensive terms $R_{tot}$ and $S_{tot}$, respectively.

\begin{equation}
JVS_{tot}=\sqrt{R_{tot}\cdot S_{tot}}
\label{eq:JVS_tot}
\end{equation}
This formulation extends the original JVS by considering not only whether the prediction remains correct under perturbations, but also how confidently the prediction is produced. As a consequence, configurations that preserve both classification correctness and large confidence margins achieve lower scores, providing a more comprehensive assessment of the overall vulnerability of Split Computing systems. Accounting for both Accuracy degradation and Confidence score degradation. \footnote{The implementation used for the combined fault-injection and adversarial-attack experiments is available in this link: \url{https://github.com/Reliable-and-Secure-Systems/DFTS_JASPER_Security}}

\subsection{Experimental Results}

Figure~\ref{fig:jvs} reports the JVS metric for different SC configurations over a progressive level of perturbation. The definition of a joint reliability-security metric requires the introduction of a common perturbation space. Reliability experiments are characterized by the BER, whereas security experiments are parameterized by the adversarial perturbation magnitude $\epsilon$. 

Therefore, a JVS is defined to establish a unified perturbation axis. Security perturbation levels are projected onto this axis, mapping $\epsilon$ with the BER value corresponding to corruption intensity. This mapping aligns and aggregates reliability and security measurements within a common experimental space.

Figure~\ref{fig:jvs} reports the evolution of JVS under increasing perturbation levels for different Split Computing configurations. As expected, JVS consistently increases with perturbation intensity, indicating that the combined vulnerability of the system becomes progressively more severe as hardware faults and adversarial perturbations intensify. This confirms that the proposed metric effectively captures their cumulative impact within a single quantitative framework.

Moreover, the different trends observed across the evaluated configurations demonstrate that JVS is sensitive not only to the perturbation level but also to the selected Split Computing strategy. Configurations exhibiting steeper increases in JVS are more susceptible to combined reliability and security threats, whereas flatter curves indicate greater robustness over a wider perturbation range. For both the CRBQ and Entropic Student approaches, higher compression factors generally result in larger JVS values, while lower compression settings consistently maintain smaller JVS values across the evaluated perturbation range. These results demonstrate that the proposed metric can effectively distinguish alternative Split Computing configurations according to their overall robustness, providing a unified basis for evaluating the trade-off between reliability and security.

Across the $JVS_{tot}$ metric shown in Figure~\ref{fig:JVStot}, the CRBQ configurations in Figure~\ref{fig:JVStot_crbq} exhibit the highest vulnerability, particularly for CRBQ 9 and CRBQ 6. These results confirm that increasing the compression of the transmitted feature maps amplifies the combined impact of reliability and security threats on the overall Split Computing system. In contrast, the Entropic Student configurations shown in Figure~\ref{fig:JVStot_entropic} achieve their lowest vulnerability for $\beta=5.12$, indicating a more favorable trade-off between model compression and robustness.

As expected, all configurations exhibit a monotonic increase in $JVS_{tot}$ as the perturbation level increases, reflecting the growing impact of both prediction errors and confidence degradation. Compared with the original JVS metric, $JVS_{tot}$ provides a clearer separation between competing configurations by accounting for the confidence associated with model predictions in addition to classification correctness. Consequently, configurations with similar misclassification rates but different confidence margins can be effectively distinguished, providing a more comprehensive assessment of the robustness of Split Computing systems under combined reliability and security threats.

While $S_{tot}$ and $R_{tot}$ are computed under a single corruption axis ($\epsilon$ or BER) varied in isolation, considering a specific threat, the  $JVS_{tot}$ stresses the model with $\epsilon$ and BER \emph{jointly}, cosidering the two threats, revealing a compounding fragility that the marginal metrics miss entirely: e.g.\ for CRBQ 1, $S_{tot}$ and $R_{tot}$ remain essentially unchanged between scale $0.01$ and $0.03$, yet $JVS_{tot}$ exposing a sharp low-scale collapse under simultaneous corruption that single-axis sweeps cannot detect. This drop is most severe precisely for CRBQ $\le 3$, $\beta\!\ge\!2.56$, suggesting their apparent robustness in the $S_{tot}$/$R_{tot}$ plots is an artefact of testing each perturbation mode in isolation rather than under realistic, co-occurring noise.
These findings highlight the importance of the proposed $JVS_{tot}$ metric in capturing the interaction between hardware faults and adversarial attacks, enabling a more comprehensive assessment of Split Computing robustness than independent reliability or security analyses.

Furthermore, $JVS_{tot}$ can inform the selection of a split computing configuration. For example, for the Entropic Student, looking at Beta = 2.56 for low BER, it seems that from the reliability $R_{tot}$ point of view \ref{fig:Rtot_entropic} this is one of the best configurations, while considering $JVS_{tot}$, Beta = 2.56 is actually one of the worst configurations. A similar effect occurs if we consider only $S_{tot}$ on the other side: in this case, the best configuration is Beta = 5.12 \ref{fig:Stot_entropic}, regardless of $\epsilon$. However, if we look at $JVS_{tot}$, this is only partially true. Beta = 5.12 remains the best for medium/high values of $\epsilon$, but for low values of $\epsilon$ the best configuration is Beta = 0.64.
This case has real design implications, showing that $JVS_{tot}$ informs the selection or modification of a split computing configuration better than $R_{tot}$ and $S_{tot}$ taken individually.

\section{Conclusion}
This paper presents a unified framework for jointly assessing reliability and security in Split Computing. First, reliability is characterized through neuron-level fault injection using the Mean Relative Accuracy Degradation (MRAD), while security is characterized through feature-map-aware adversarial attacks simulations using the Attack Success Rate (ASR). Based on these complementary analyses, the Joint Vulnerability Score (JVS) is introduced, along with a confidence-aware extension that jointly captures prediction errors and confidence degradation. The framework is evaluated on ten Split Computing configurations based on ResNet-50 trained on ILSVRC-2012. Experimental results show substantial differences across compression strategies, with MRAD ranging from 44.3\% to 61.2\% under fault injection, while adversarial attacks achieve up to 98.8\% ASR. Furthermore, the proposed joint metrics reveal vulnerability trends that remain hidden when reliability and security are analyzed independently, providing a more comprehensive methodology for designing dependable Split Computing systems.

\section*{ACKNOWLEDGMENT}
\small
This work was supported in part by the Estonian Research Council grant PUT PRG1467 ``CRASHLESS'', EU Grant Project 101160182 ``TAICHIP'', and by the Federal Ministry of Research, Technology and Space of Germany (BMFTR) for supporting Edge-Cloud AI for DIstributed Sensing and COmputing (AI-DISCO) project (Project-ID ``16ME1127'').

\bibliographystyle{IEEEtran}
\bibliography{ref.bib}

\typeout{get arXiv to do 4 passes: Label(s) may have changed. Rerun}
\end{document}